\documentclass[pdflatex,sn-mathphys-num]{sn-jnl}

\usepackage{graphicx}%
\usepackage{multirow}%
\usepackage{amsmath,amssymb,amsfonts}%
\usepackage{amsthm}%
\usepackage{mathrsfs}%
\usepackage[title]{appendix}%
\usepackage{xcolor}%
\usepackage{textcomp}%
\usepackage{manyfoot}%
\usepackage{booktabs}%
\usepackage{algorithm}%
\usepackage{algorithmicx}%
\usepackage{algpseudocode}%
\usepackage{listings}%
\usepackage{epstopdf}%
\usepackage{svg}
\usepackage{amsmath}
\usepackage{xcolor}
\usepackage{ulem}
\usepackage{url}

\begin{document}

\title[]{On the relationship between noise squeezing and Rabi oscillations in active quantum dot ensembles}

\author[1]{\fnm{Ori} \sur{Gabai}}
\equalcont{These authors contributed equally to this work.}

\author[1]{\fnm{Amnon} \sur{Willinger}}
\equalcont{These authors contributed equally to this work.}

\author[1]{\fnm{Igor} \sur{Khanonkin}}
\equalcont{These authors contributed equally to this work.}

\author[2]{\fnm{Vitalii} \sur{Sichkovskyi}}

\author[2]{\fnm{Johann Peter} \sur{Reithmaier}}

\author*[1]{\fnm{Gadi} \sur{Eisenstein}}\email{gad@technion.ac.il}

\affil[1]{\orgdiv{Faculty of Electrical and Computer Engineering and the Helen Diller Quantum Center}, \orgname{Technion - Israel Institute of Technology}, \orgaddress{\city{Haifa}, \country{Israel}}}

\affil[2]{\orgdiv{Institute of Nanostructure Technologies and Analytics, Technische Physik, CINSaT}, \orgname{Kassel University}, \orgaddress{\city{Kassel}, \country{Germany}}}

\abstract{Squeezed light is usually generated using passive nonlinear materials. Semiconductor lasers and optical amplifiers (SOAs) also offer nonlinearities but they differ in that they add amplified spontaneous emission (ASE). Squeezing to below the vacuum level has been demonstrated in a semiconductor laser, and gain saturation in SOAs can likewise reduce photon-number fluctuations to, and in some cases below, the vacuum limit. Here, we demonstrate that Rabi oscillations in room-temperature quantum-dot SOAs, induced by short resonant pulses, cause cyclical noise modification that repeat with every change of 2$\pi$ in pulse area, corresponding to a fourfold increase in excitation pulse energy. Homodyne measurements reveal in those cases  elliptical Wigner functions corresponding to squeezed thermal states and in certain regimes, the state is squeezed to below the vacuum level. At other pulse areas, the Wigner functions are circular representing thermal coherent states. This periodic behavior persists over two orders of magnitude in input pulse energy, spanning several 2$\pi$ cycles. Under specific bias and excitation conditions, we further observe a non-Gaussian Wigner function featuring two bright lobes. Although its precise nature remains unresolved, this structure may be consistent with a Schrödinger cat–like state whose accompanying negativity is suppressed due to an approximately 10 dB optical output loss. Notably, the emergence of this non-Gaussian state is itself periodic in excitation pulse energy.}


\maketitle

\section{Introduction}\label{sec1}
Squeezed light is a non-classical state of an optical signal in which the uncertainty in one quadrature (for example, the photon number) is reduced at the expense of the other quadrature (the phase), while obeying the Heisenberg uncertainty~\cite{PhysRevD.23.1693,PhysRevLett.55.2409,andersen_16}. Squeezed optical signals allow for improved fiber communication~\cite{slusher2002squeezed,yuen2004communication,suleiman202240}, and are imperative for precise measurements in quantum sensing and metrology~\cite{aasi_13}, as well as for continuous variable quantum communication~\cite{andersen_15,takeda_19}. Generation of squeezed light requires a strong nonlinearity which is most often implemented in passive $\chi^{(2)}$ or $\chi^{(3)}$ materials~\cite{boyd_23,wu1986generation,mccormick2006strong,heersink2005efficient,bergman1991squeezing}.

Strong nonlinearities are also available in optically active media, such as semiconductor lasers and optical amplifiers (SOAs). Active media differ from passive materials in that they add amplified spontaneous emission noise (ASE). The nonlinearity in an active device, and its consequent ability to squeeze the noise, stems from saturation which is determined, in turn, by the medium dynamical properties. The dynamics of semiconductor devices are fast, and this has a major impact on the noise properties under saturated conditions \cite{bilenca2004noise,berg2004saturation,bogoni2004modeling}. 

Squeezing in a DC driven semiconductor laser was demonstrated by Y. Yamamoto's group several decades ago~\cite{yamamoto1992photon,machida1987observation}. Similar squeezing levels were  achieved in A. Yariv's group by applying a weak dispersive feedback to the laser~\cite{kitching1995room}. A recent paper~\cite{zhao2024broadband} reported squeezing in a quantum dot (QD) laser driven by an ultra-low-noise current source. A model for squeezing in semiconductor lasers was described in~\cite{vey1997semiclassical}.
The noise of a saturated SOA was shown both theoretically and experimentally to exhibit a spectral hole in the vicinity of the saturating signal~\cite{shtaif1996noise,bilenca2004noise, capua2008direct}. Clear narrowing of the the statistical distribution of the power spectral density in a saturated SOA was demonstrated experimentally for continuous waves (CW)~\cite{shtaif2002experimental} as well as pulse~\cite{bilenca2005statistical} excitations, and confirmed by a simulation~\cite{bilenca2005fokker}. 

Ultra short pulse propagation in an SOA induces several nonlinear coherent interactions~\cite{eisenstein2019coherent}. Once the pulses are sufficiently intense and their duration is short in comparison to the semiconductor coherence time (which is roughly 1 ps at room temperature~\cite{khanonkin2021room}), they induce coherent interactions such as Rabi oscillations, which were demonstrated experimentally mainly in QD SOAs~\cite{eisenstein2019coherent,capua2014rabi,borri2002coherent,kolarczik2013quantum}, but also in a quantum cascade laser~\cite{choi2010ultrafast}, and were modeled for quantum well SOAs~\cite{zhang2004rabi}. This means that during the pulse, the SOA oscillates between gain and absorption, and this leaves a direct imprint on the temporal profiles of the amplitude and phase of the SOA output pulse. The carrier density in the SOA also oscillates and this affects the amplified spontaneous emission noise accompanying the pulse.

We demonstrate here, for the first time, a direct interplay between Rabi oscillations and noise in a QD SOA operating at 1550 nm and at room temperature. To induce Rabi oscillations, we excite the QD SOA with resonant 100 fs pulses and characterize the noise of the resulting output pulses using homodyne detection, where a replica of the excitation pulse acts as the local oscillator (LO). During the portions of the Rabi cycle when the quantum dots provide gain, the uncertainty in the photon number is reduced yielding an elliptical Wigner function. We highlight three consequent results. First, the noise modification is cyclical with the pulse excitation energy repeating with every fourfold increase of the pulse energy which amounts to a 2$\pi$ increase in pulse area namely, a full cycle on the Bloch sphere. Some of these elliptical Wigner functions represent squeezed states and others describe squeezed coherent thermal states~\cite{kim1989properties}. Other pulses areas yield circular Wigner functions describing coherent thermal states~\cite{mann1989thermal} resulting from the addition of the SOA ASE.
This periodic behavior extends over a very wide range of excitation pulse energies, from 1.58 pJ to 158 pJ. Second, under specific conditions, the photon number uncertainty is clearly squeezed to below the vacuum level. Finally, in some cases, the obtained Wigner function is non-Gaussian comprising two separated bright lobes. The nature of this Wigner function is unclear as it does not represent any known quantum state. It may represent however, a Schr\"odinger cat-like state where the corresponding negativity is not observable due to a 10 dB output optical loss. The observation of this non-Gaussian state is also cyclical, recurring with a 2$\pi$ increase in pulse area.
 
\section{Results and Discussion}\label{sec2}

\begin{figure}[ht]
    \centering
    \includegraphics[width=0.75\linewidth]{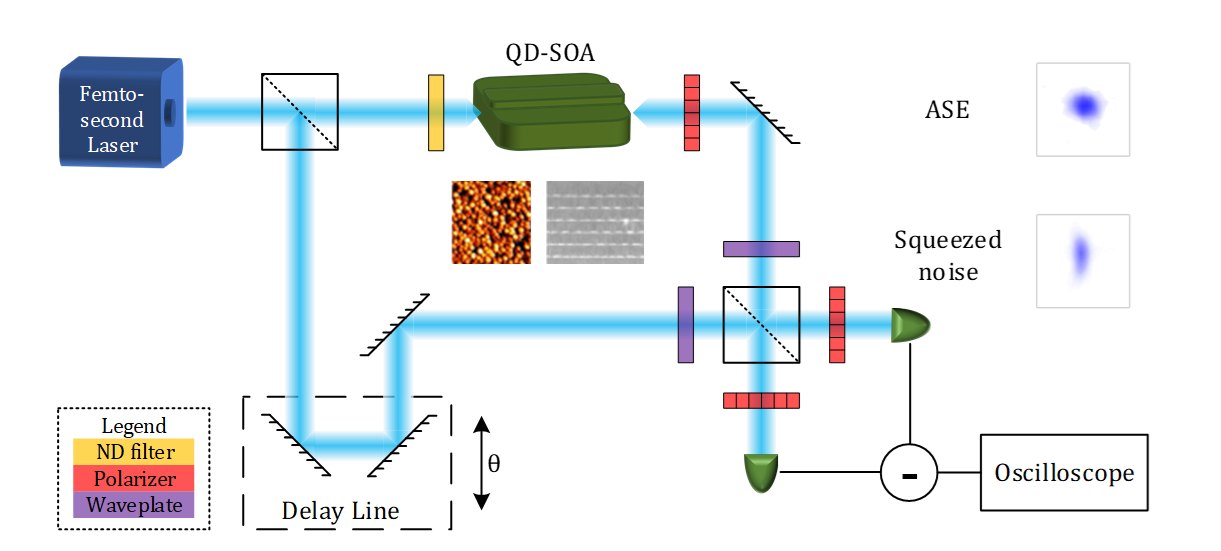}
    \caption{Experimental setup for homodyne detection of femtosecond pulses following the interaction with a QD ensemble. The two small insets show Wigner functions of the spontaneous emission noise under linear conditions (no input signal) and of a squeezed state.}
    \label{fig_setup}
\end{figure}

The experimental setup for homodyne characterization of ultra short pulses following coherent interaction with a room-temperature QD SOA is shown in Fig. \ref{fig_setup}. A 100 fs laser pulse is split with one part serving as the LO and the second is coupled to a QD optical amplifier. Following the interaction with the QD SOA, the transmitted field (signal) is recombined with the LO, whose relative phase is controlled via an optical delay line. The resulting interference is detected by a balanced photodetector, and the difference signal is recorded using an oscilloscope.
Scanning the LO phase over 2$\pi$ radians (a periodic motion over $2~\mu m$) for about one thousand times yields a histogram of the current uncertainty. The histogram is transformed into a Wigner function using the inverse Radon transform~\cite{deans2007radon}. The insets in Fig. \ref{fig_setup} show respectively, the Wigner functions of the ASE noise under linear conditions namely, with no input signal and of a typical squeezed state.  

\begin{figure}[bp!]
    \centering
    \includegraphics[width=1\linewidth]{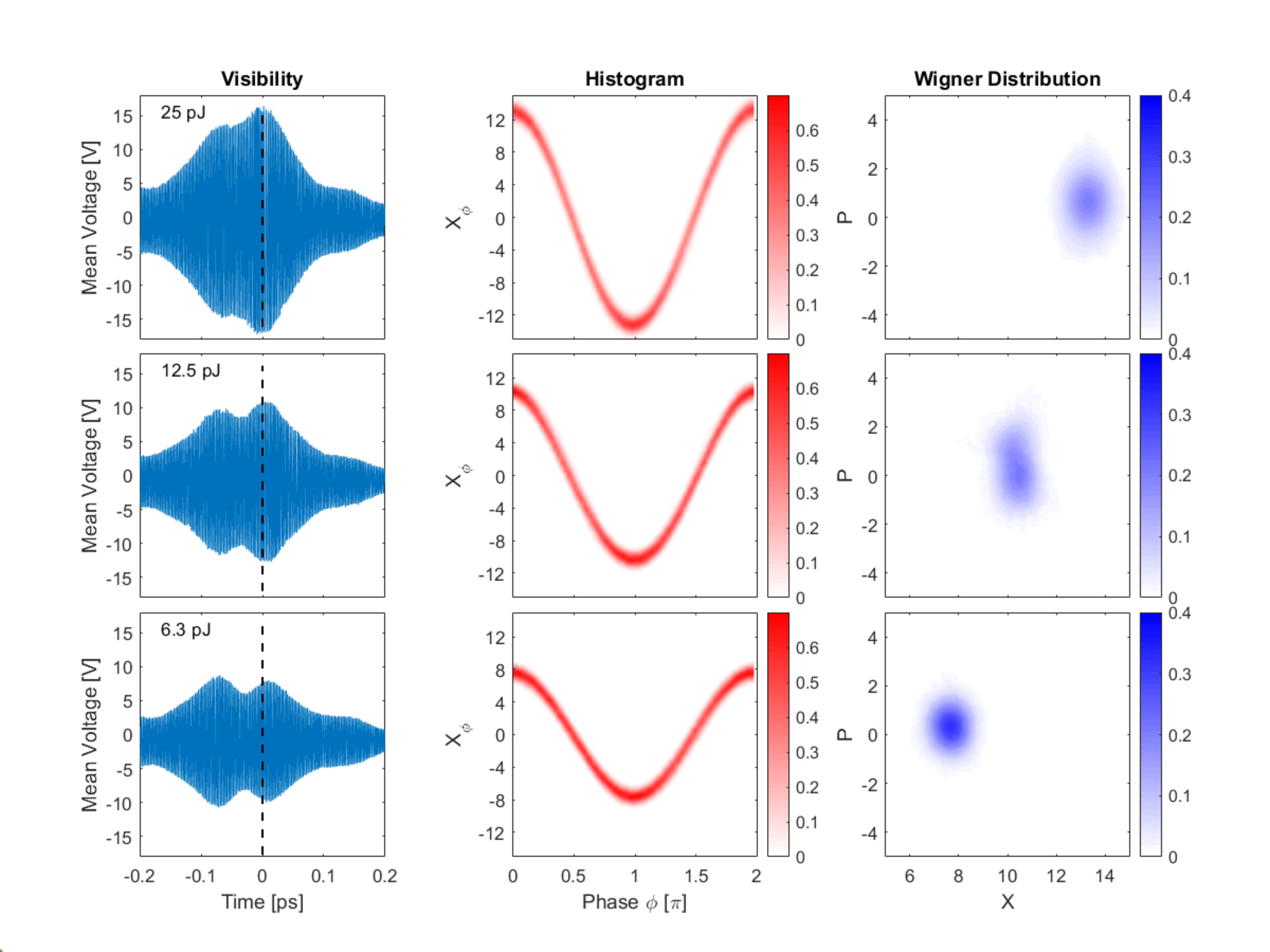}
    \caption{\textbf{Measurements of the homodyne system for three different excitation pulse energies}. The left column represents the measured temporal correlation between the LO and the signal. The middle column represents the histogram of the normalized output current of the balanced photodetector as a function of the LO phase measured at 0 ps, i.e., the trailing peak, which is indicated by the dashed line in the correlation graph. The right column describes the corresponding Wigner functions. X and P are the quadratures for which $X^2+P^2$ equals the square of the photon number.}
    \label{fig_wigner_set1}
\end{figure}

The homodyne system serves also as a cross-correlator where the LO pulse samples the signal pulses after propagating through the QD gain medium while the relative temporal delay between them is scanned. Such cross-correlations for three different excitation pulse energies are shown in the left column of Fig. \ref{fig_wigner_set1}. The cross-correlations reveal that the 100 fs excitation pulse broadens to about 1 ps and attains an amplitude profile resembling the results measured using cross-correlation frequency resolved optical gating (X-FROG) \cite{capua2014rabi}.
Homodyne measurements were performed with the LO sampling the signal at different temporal locations. The center column of Fig. \ref{fig_wigner_set1} describes histograms of the normalized output current of the balanced photodetector as a function of the LO phase, measured at the trailing peak (dashed line at 0 ps). The corresponding extracted Wigner function are depicted in the right column, revealing a clear dependence on the input pulse energy. At 6.3 and 25 pJ, the circular Wigner functions represent thermal coherent states. However, at 12.5 pJ it exhibits a clear elliptical function characteristic of the reduced photon number uncertainty. This is a squeezed coherent thermal state. At any other temporal position, the Wigner functions for all excitation pulse energies exhibit coherent states, namely, they are not affected by the QD dynamics stemming from the Rabi oscillations.

\begin{figure}[ht]
    \centering
    \includegraphics[width=1\linewidth]{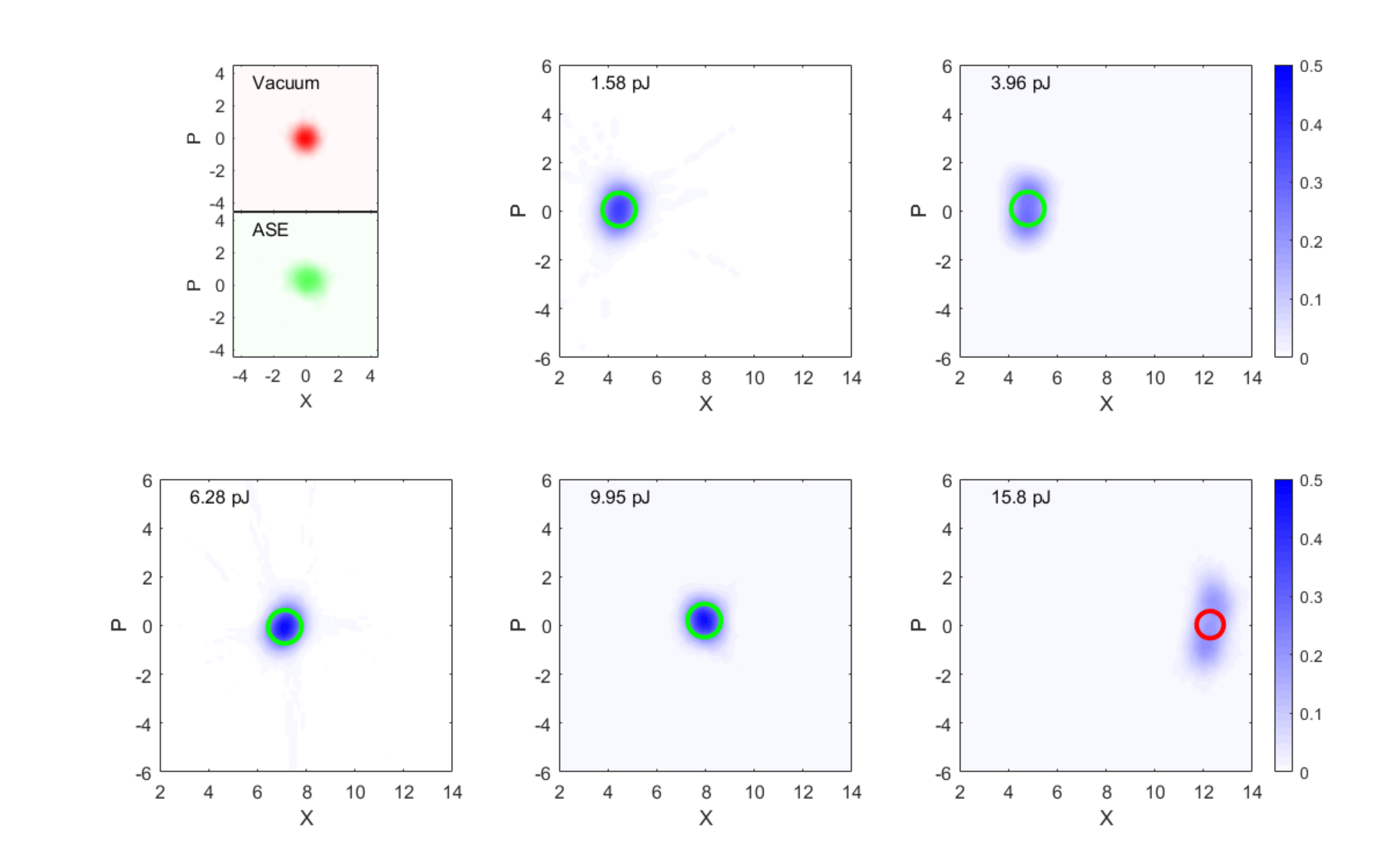}
    \caption{\textbf{Excitation energy dependent Wigner functions}. The left panel shows the Wigner distributions of the vacuum (red) and the ASE (green). At 1.56 pJ, 6.28 pJ and 9.95 pJ, the Wigner functions are round, representing coherent thermal states. Each is compared to the standard deviation of the ASE (green circle) and are found to be equal or slightly wider than the ASE. The Wigner functions at 3.96 pJ and 15.8 pJ are elliptical. $\Delta X$ for the 3.96 pJ case is equal to the width of the ASE. However, at an excitation energy four times large namely at 15.8 pJ, $\Delta X$ is slightly smaller than the standard deviation of the vacuum shown in a red circle and hence represents a squeezed state.}
    \label{fig_wigner_set2}
\end{figure}

The dependence on pulse excitation energy was examined in detail with the results shown in Fig. \ref{fig_wigner_set2}. The upper left panel shows for reference, the Wigner distributions of the vacuum and the ASE. Three Wigner functions, measured at 1.56 pJ, 6.28 pJ and 9.95 pJ, are round describing coherent thermal states. Their distribution is equal or is slightly wider than the standard deviation of the ASE which is shown in a green circle. The Wigner functions for 3.96 pJ and for 15.8 pJ (which is four ties larger) are elliptical. $\Delta X$ for the case of 3.96 pJ is equal to the standard deviation of the ASE, namely this is a squeezed thermal state. Increasing the input energy by a factor of four to 15.8 pJ yields a Wigner function whose uncertainty in the photon number is slightly, but nevertheless clearly, smaller than the standard deviation of the vacuum (shown in a red circle). It represents squeezing.

\begin{figure}[ht]
    \centering
    \includegraphics[width=1\linewidth]{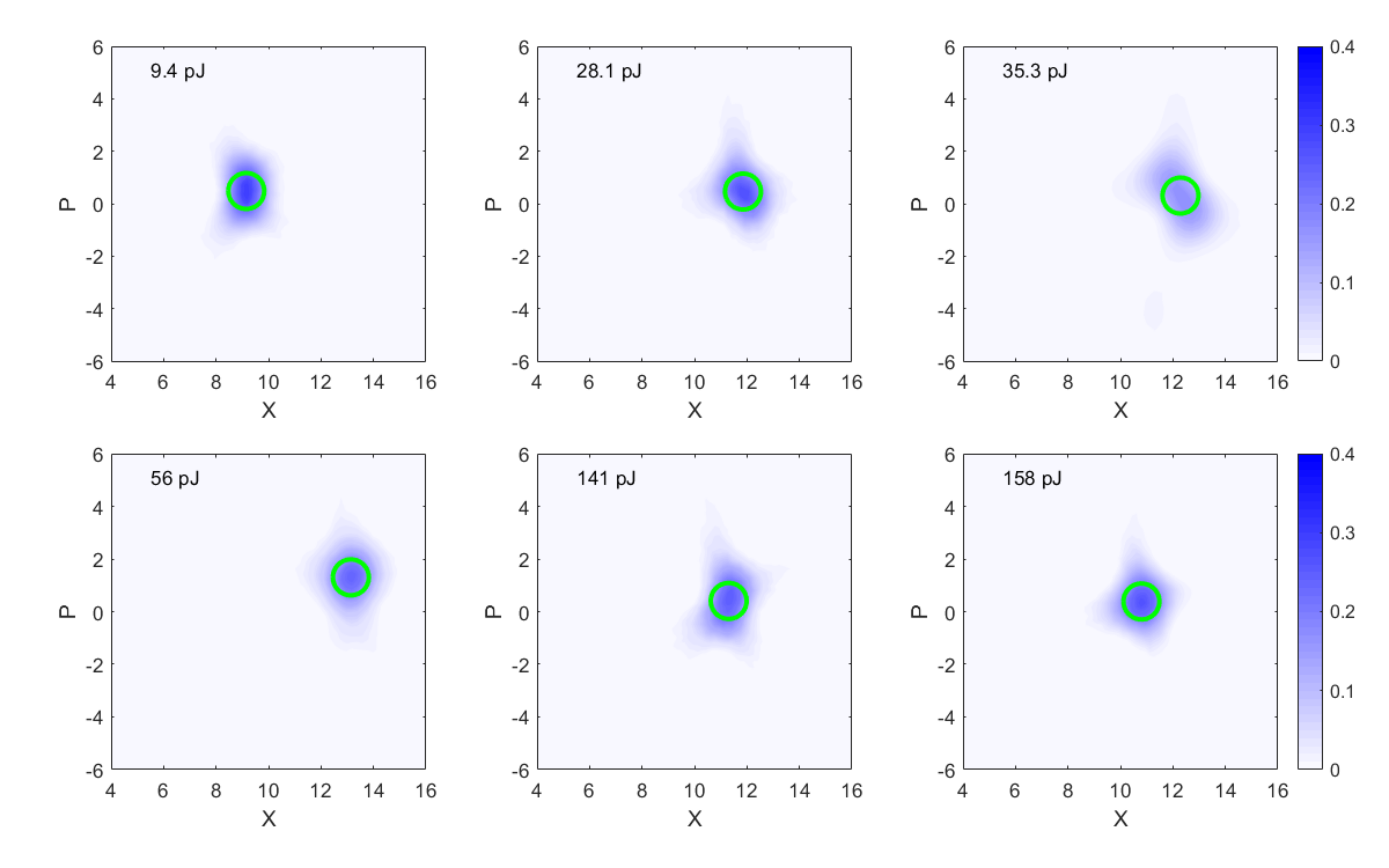}
    \caption{\textbf{Wigner functions for large excitation energies} At 9.4 pJ, 35.3 pJ  and 141 pJ, the Wigner functions are elliptical while all other energies, yield round Wigner functions representing thermal coherent states. These are equal or slightly larger than the uncertainty of the ASE, shown as green circles. The elliptical Wigner functions at 9.4 pJ 35.3 pJ and 141 pJ are squeezed thermal states with their $\Delta X$ being similar to the ASE shown in green circles.}
    \label{fig_wigner_set3}
\end{figure}

The experiments were repeated at a range of larger input energies, from 9.4 pJ to 158 pJ, under conditions of somewhat lower input coupling efficiencies to the QD SOA. The results are shown in Fig. \ref{fig_wigner_set3}. 
 The cyclical nature, with the same fourfold periodicity of the excitation energies is maintained for 9.4 pJ, 35.3 pJ and 141 pJ, yielding elliptical Wigner functions. $\Delta X$ in those is similar to the standard deviation of the ASE distribution (shown in green circles). Therefore, they do not represent squeezing but rather squeezed coherent thermal states. The higher energies yield multiple Rabi cycles, each contributing a different degree of saturation and hence, the cumulative effect of the noise modification is less efficient. The Wigner functions at 28.1 pJ, 56 pJ and 158 pJ are circular and describe once more coherent thermal states. Some elliptical Wigner functions in Fig. \ref{fig_wigner_set2} and Fig. \ref{fig_wigner_set3}, in particular for the higher excitation energies, are tilted. This means that in addition to the obvious relationship between photon number and phase uncertainties, there exists a nonlinear coupling between them. This is a consequence of the power dependence of the Henry $\alpha$ parameter~\cite{septon2019large,bjelica2016optimization}.
 
\begin{figure}[htp!]
    \centering
    \includegraphics[width=0.75\linewidth]{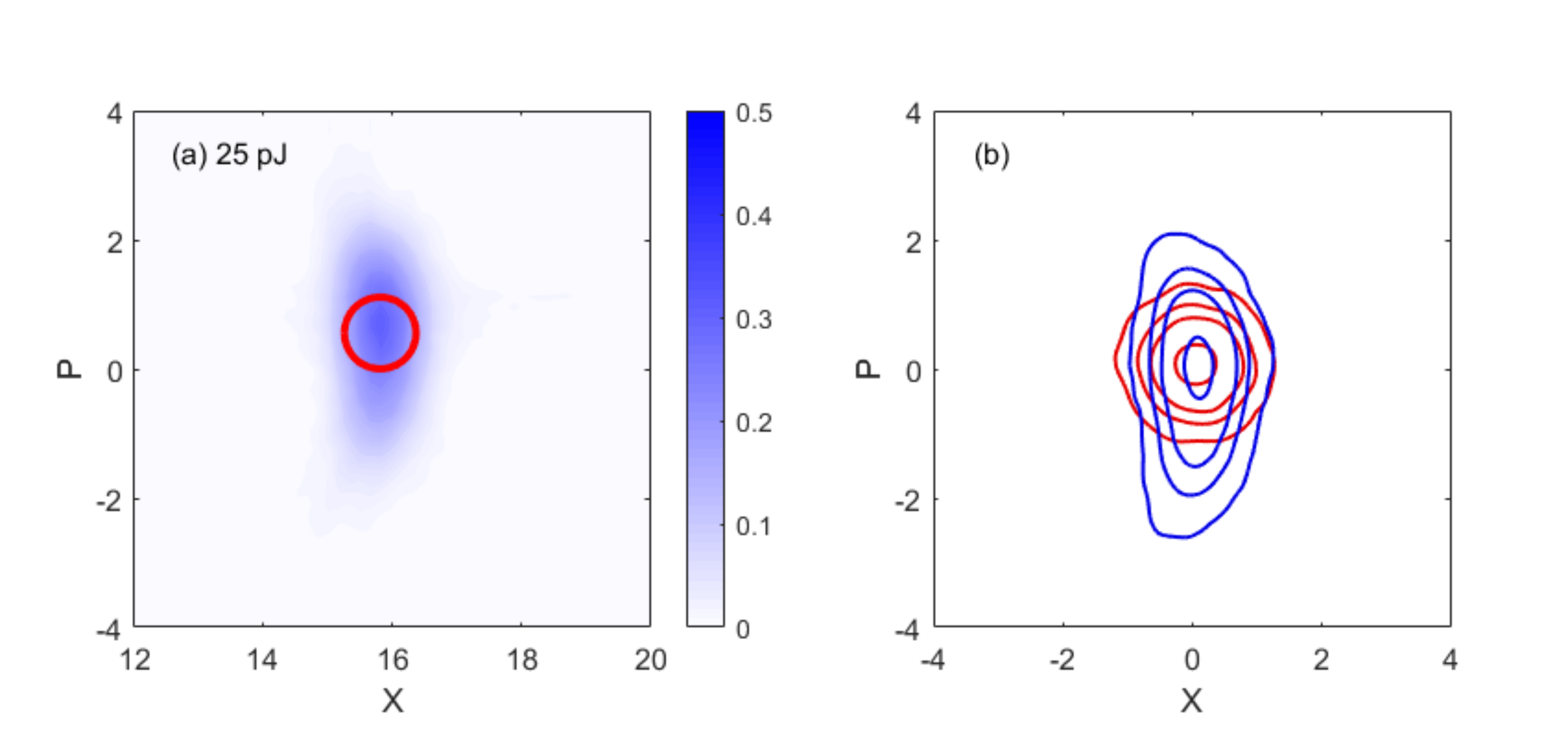}
    \caption{\textbf{Squeezed state at an excitation of 25 pJ} (a) Squeezed Wigner function distribution. The red circle represents the vacuum. (b) contour plots of the Wigner and vacuum distributions highlighting the squeezed state. The full width at half maximum is reduced from 1.43 for the vacuum state to 1.17 for the squeezed state.}
    \label{fig_wigner_and_contour}
\end{figure}

\begin{figure}[htp!]
    \centering
    \includegraphics[width=0.75\linewidth]{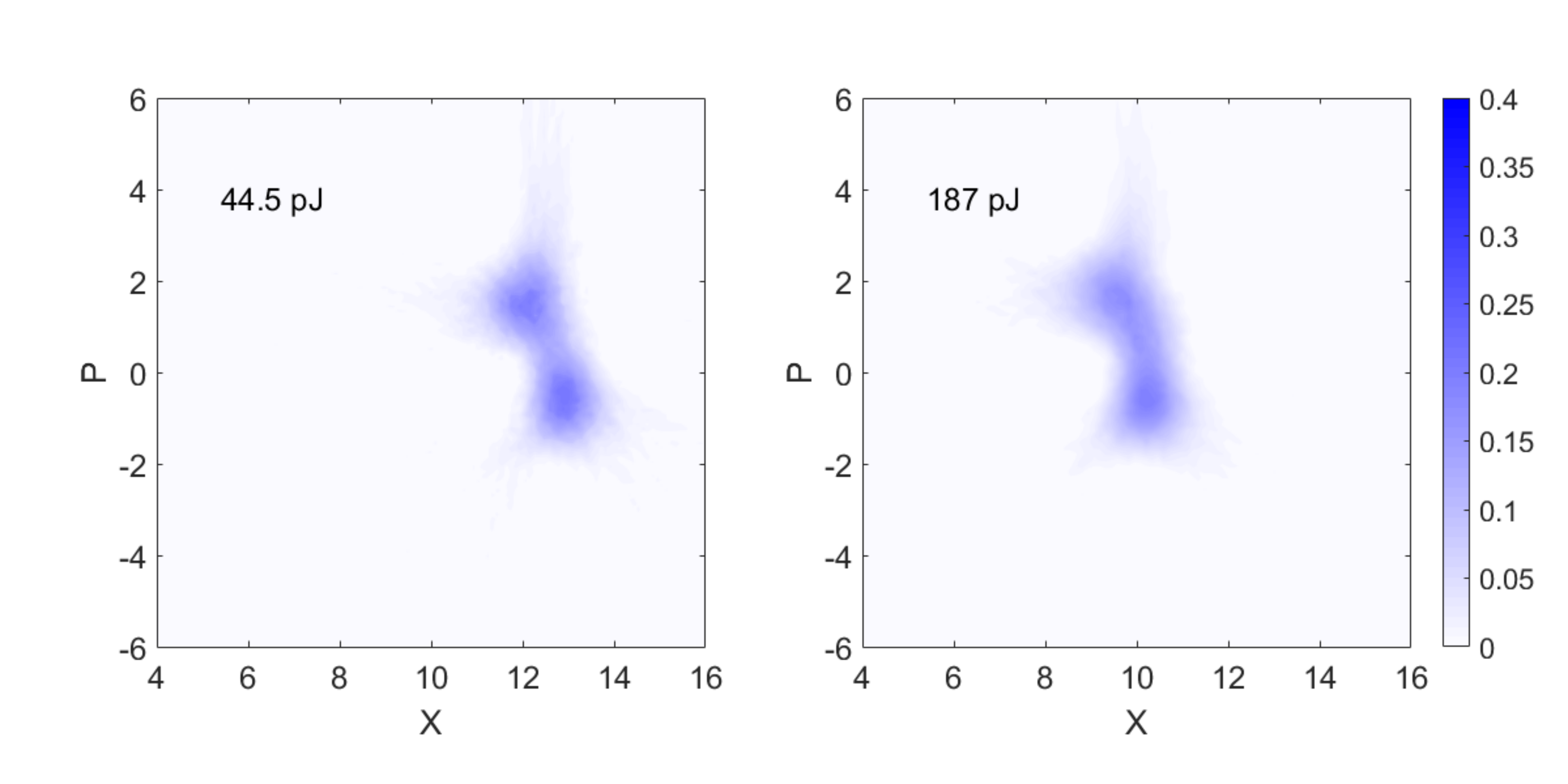}
    \caption{\textbf{Non-Gaussian Wigner functions obtained for 44.5 pJ and 187 pJ (a fourfold increase).} This Wigner function does not represent any known quantum state. However, it resembles a Schr\"odinger cat-like state except that the negativity between the two bright lobes is absent, possibly due to the system optical losses of 10 dB.}
    \label{fig_wigner_set4}
\end{figure}

 The QD SOA is a distributed, inhomogeneously broadened medium where an accurate value for the pulse area is hard to define even though the differences in area between successive measurements are essentially accurate. It turns out that there are some specific excitation energies that yield very unique results. The first is shown in Fig. \ref{fig_wigner_and_contour}. For an excitation energy of 25 pJ, the obtained Wigner function is unequivocally squeezed below the vacuum, as seen in Fig. \ref{fig_wigner_and_contour} (a). Fig. \ref{fig_wigner_and_contour} (b) shows a contour plot of the Wigner (in blue) and vacuum (in red) distributions. The state is clearly squeezed.
Finally, under excitation energies of 44.5 pJ and 187 pJ, the obtained Wigner function is non-Gaussian as seen in Fig. \ref{fig_wigner_set4}. This Wigner function that comprises two bright lobes does not represent any known quantum state. It does resemble though a Schr\"odinger cat-like state except that their corresponding negativity is not observed. The negativity is possibly missing due to a large, close to 10 dB output optical loss~\cite{leonhardt1997measuring} .

\section{Conclusion}
In this work, we demonstrated a direct interplay between
Rabi oscillations in a room-temperature QD SOA, induced by an ultra-short excitation pulse and noise. Specifically, Rabi oscillations cause
cyclical modifications of the output photon number uncertainty. This effect repeats with every fourfold increase in the excitation pulse energy, equivalent to a 2$\pi$ increment in pulse area. The noise in those cases takes on a value which is similar or slightly smaller than the ASE level of a linear amplifier. In some cases, it is actually squeezed to below the vacuum level. This behavior was observed for a two orders of magnitude range of excitation energies. Naturally, the conjugate phase quadrature noise increases correspondingly.

The underlying mechanism arises from the periodic evolution between gain and absorption in the QD SOA during Rabi oscillations. The photon number uncertainty is reduced during portions of the Rabi cycle when the QDs provide gain saturation; otherwise, the output exhibits the noise properties of a coherent thermal state. However, there are some specific excitation levels that yield very unique results. A clear squeezed state and a non-Gaussian Wigner functions were demonstrated. The latter resembles a Schr\"odinger  cat-like state where the negativity is omitted, possibly due to a large optical loss between the SOA output facet and the homodyne system.

Our findings represent a clear manipulation of noise (amplifier added ASE and noise carried by the input signal) via coherent light–matter interactions. This is a fundamentally new regime of amplifier operation that could enable engineered noise profiles in active photonic devices. Potential applications include tailored noise suppression in optical amplifiers for quantum communication systems, or enhanced sensitivity in measurement schemes where reduced amplitude noise, is crucial.

Future research should explore this effect in cavity-based system with tailored feedback. Additionally, incorporating pulse shaping techniques to precisely control and optimize the temporal dynamics of the Rabi oscillations \cite{karni2016coherent} may enhance the noise reduction while opening new avenues for deterministic quantum noise engineering on chip-scale photonic platforms.

\section{Methods}\label{sec11}

\subsection{Experimental setup}
The experimental setup consists of a femtosecond fiber-based laser (Toptica FemtoFiber pro) operating at a center wavelength of 1550 nm, delivering pulses with a duration of approximately 100 fs. The optical path length was controlled using a motorized optical delay line (Physik Instrumente, model V-508, step resolution of 2 nm) to achieve precise temporal scanning. The output beams were combined and directed to a balanced photodetector (Femto, model HBPR-100M-60K-IN-FST) for detection of the interference signal. The resulting photocurrent was recorded using an oscilloscope (Keysight, model EXR404A, with 4 GHz bandwidth) for subsequent analysis. The delay line was moving periodically at rate of 5 Hz and an amplitude of 2 $\mu$m and correspondingly produced a position signal that was sampled by the oscilloscope in synchronous with the signal from the balanced photodetector.
\subsection{QD SOA fabrication}
The active region of the gain medium we used comprised six layers of high-density, $6 \times 10^{10}$cm$^{-2}$, InAs QDs grown by molecular beam epitaxy in the Stranski - Krastanow mode. The optical amplifier is fabricated as a 2$~\mu\text{m}$ wide, 1.5 mm long ridge waveguide whose end facets were anti-reflection coated. The QDs exhibit a narrow photo-luminescence linewidth, measured at 10 K,  of 17 meV for a single layer and 26 meV for the six-layer stack. This linewidth represents a record uniformity compared to all reported QDs in any material system. Details of the device fabrication and characterization are given in~\cite{banyoudeh2015high}. The emission power of the measured electro-luminescence spectra increases with applied bias but the spectral shape remains unchanged~\cite{khanonkin2017ultra}.
\subsection{Wigner function analysis}
Histograms of the current at the output of the balanced photodetector, measured with a real time oscilloscope,  is constructed from one thousand, periodic scans of the delay line with an amplitude of 2~$\mu$m. Wigner functions are obtained from those histograms using the inverse Radon transform. This transform is intended for CW or narrow-spectrum signal. The spectral width of the 100 fs excitation pulses is roughly 3\% of the optical frequency and hence, the assumption of narrow band signal is valid.
\backmatter

\bmhead{Acknowledgements}
The research was partially supported by the Israel Science Foundation under grant number 460/21 and by the Hellen Diller Quantum Center at Technion. IK acknowledges financial support from the Ministry of Innovation, Science and Technology of Israel.

\end{document}